\documentclass[aps,prl,reprint]{revtex4-1}
\usepackage[english]{babel}
\usepackage{graphicx}
\usepackage{graphics}
\begin{document}


\title{Ultrahigh and persistent optical depths of caesium in Kagom\'e-type hollow-core photonic crystal fibres}

\author{Krzysztof T. Kaczmarek}
\author{Dylan J. Saunders}
\author{Michael R. Sprague}
\author{W. Steven Kolthammer}
\author{Amir Feizpour}
\author{Patrick M. Ledingham}
\author{Benjamin Brecht}
\author{Eilon Poem}
\author{Ian A. Walmsley}
\author{Joshua Nunn}
\affiliation{Clarendon Laboratory, University of Oxford, Parks Road, Oxford OX1 3PU, UK}

\begin{abstract}

Alkali-filled hollow-core fibres are a promising medium for investigating light-matter interactions, especially at the single-photon level, due to the tight confinement of light and high optical depths achievable by light-induced atomic desorption. However, until now these large optical depths could only be generated for seconds at most once per day, severely limiting the practicality of the technology. Here we report the generation of highest observed transient ($>10^5$ for up to a minute) and highest observed persistent ($>2000$ for hours) optical depths of alkali vapours in a light-guiding geometry to date, using a caesium-filled Kagom\'e-type hollow-core photonic crystal fibre. Our results pave the way to light-matter interaction experiments in confined geometries requiring long operation times and large atomic number densities, such as generation of single-photon-level nonlinearities and development of single photon quantum memories.

\end{abstract}



\maketitle

Optical quantum memories are a crucial component of photonic quantum technologies, enabling the storage and manipulation of optically encoded quantum information \cite{Humphreys2014} or temporal multiplexing of probabilistic operations \cite{Nunn2013}. Such memories require strong light-matter coupling that facilitates the transformation of a photon into a matter excitation. The coupling strength can be increased from the material side, by increasing the number of atoms interacting with light, i.e. the optical depth (OD), or from the photonic side, by confining the photons in a waveguide geometry, which overcomes the focused light interaction length limit imposed by free-space diffraction. Warm vapours contained in hollow-core photonic crystal fibres (HC-PCF) are therefore a promising candidate for such applications \cite{Benabid2002}. Another exciting approach are cold atomic gases coupled to tapered optical fibres \cite{Ayrin2015,Gouraud2015}.

HC-PCFs filled with alkali vapours have been used to demonstrate the basic building blocks required for photonic quantum information processing, i.e. single-photon level optical nonlinearities \cite{Venkataraman2013} (with the potential of using Rydberg atoms \cite{Epple2014}) and optical memories \cite{Sprague2014}. The latter application especially requires fibres of relatively large core sizes ($>10\:\mu$m), such as the Kagom\'e fibres used in this work, as the transit time of atoms through the core can be a fundamental limit to the memory efficiency and lifetime. An exciting solution to this is the use of spin-preserving coatings, but their use in confined geometries is still under investigation \cite{Bhagwat2009}.

Due to the alkali's reactivity, adsorption to the core surfaces significantly limits the ambient optical depths achievable in HC-PCFs. Two ways to overcome this limitation is to use less reactive non-alkali vapours such as mercury \cite{Vogl2014} or use ultracold alkali atoms contactlessly guided within the fibre core \cite{Blatt2014}. In the case of warm alkali vapours, on the other hand, Light-Induced Atomic Desorption (LIAD) has been used to generate high transient ODs in hollow-core fibres \cite{Slepkov2008}. High-power infrared laser light stimulates the desorption of alkalis via at least two pathways: one, the direct vapourisation of metallic layers, and two, surface-plasmon enhanced heating of metallic nanoclusters \cite{Burchianti2008}. In bulk glass cells, LIAD was used to generate persistent moderate increases in optical depth (600\% over 15 minutes) \cite{Burchianti2010}. However, the persistency of high OD using LIAD has been elusive. Here, we report the observation of ultra-high transient and high persistent optical depth of caesium (Cs) confined in large-core Kagom\'e HC-PCFs of two different core sizes. We also develop a phenomenological model that agrees with our observations.

\begin{figure}[h!]
\centerline{\includegraphics[width=\columnwidth]{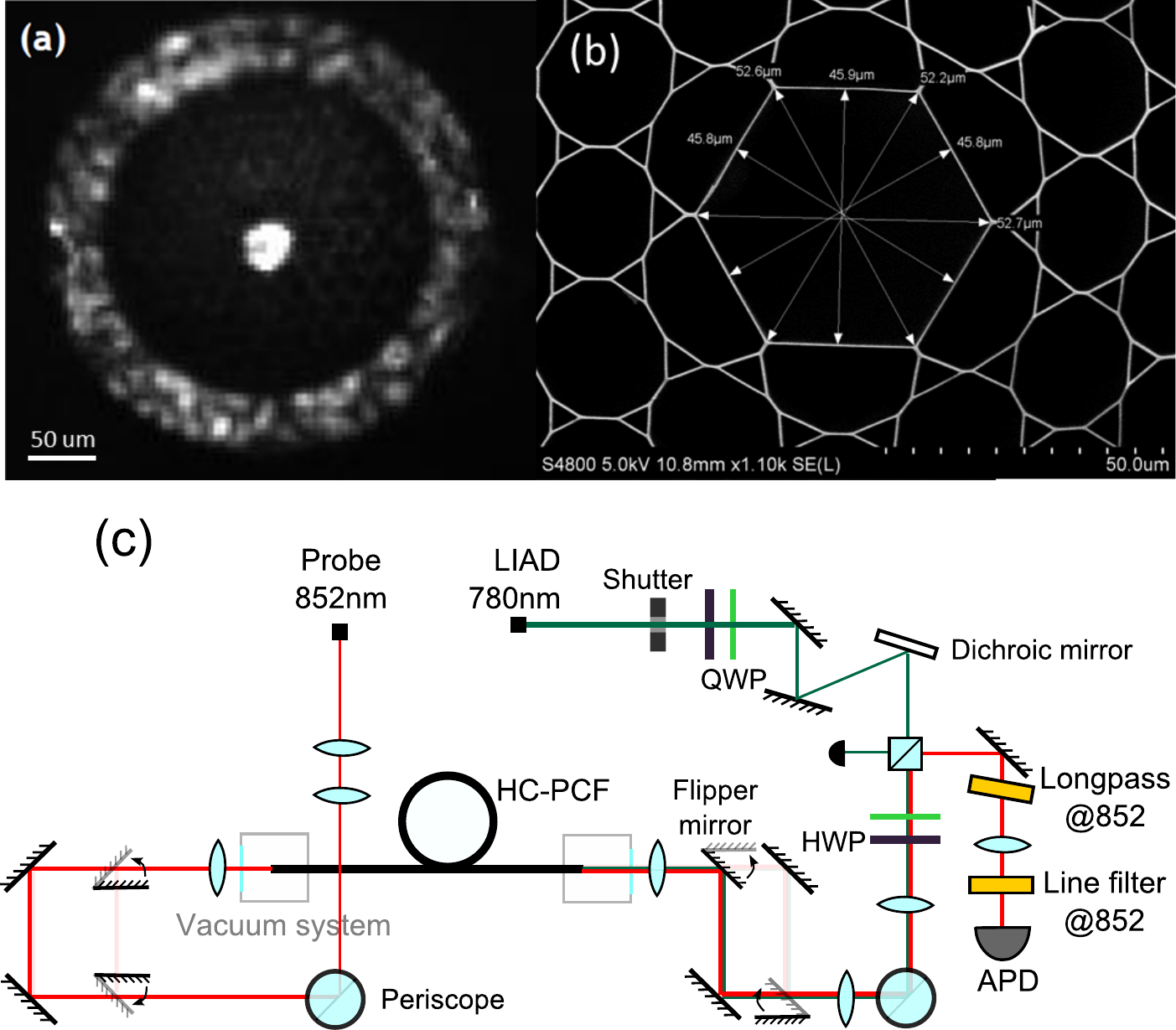}}
\caption{(a) Optical image of the end face of the $26\:\mu$m HC-PCF with light coupled into the core. (b) Scanning electron microscopy image of the $46\:\mu$m HC-PCF core. (c) Schematic of the experiment.}\label{fig:setup}
\end{figure}

The experiment was the following. We have investigated the effect of LIAD on two 20 cm Kagom\'e-structured HC-PCFs of $26\:\mu$m and $46\:\mu$m core diameters (Fig. \ref{fig:setup} (a) and (b) respectively). The two fibres were inserted inside a single vacuum chamber, which was consequently baked and evacuated to a base pressure $<10^{-9}$ Torr. The chamber was then passively loaded with caesium from an ampoule inside the system. After loading the chamber for a month at $80^o$C, we turned down the temperature to a room level of $21^o$C, keeping the windows slightly warmer in order to avoid Cs accumulation, and allowed it to further load for 12 months, over which we conducted experiments. This time could be decreased if the system was kept at a higher temperature throughout the loading period.

In order to measure the optical depth of caesium vapours inside the fibres, we send in a weak probe beam [$4 (12)$ nW into the $26 (46)\:\mu$m HC-PCF] derived from an external cavity diode laser (ECDL), that is scanned in frequency over the D2 line of Cs ($6S_{1/2}(F=3)\rightarrow6P_{3/2}$, at 852 nm) and coupled into the fibre core. We use a combination of flipper mirrors and translation stages for switchable coupling into the two fibres, as shown in Fig. \ref{fig:setup} (c). The transmitted light is subsequently measured on an avalanche photodiode (Thorlabs APD110A/M), operated in linear mode.

The resultant transmission profile is fitted to a series of Voigt functions, i.e. convolutions of the Lorenzian function due to homogenous broadening and Gaussian function from the inhomogenous broadening of the transitions from the different hyperfine states of $6P_{3/2}$. The absorption pre-factor $d^*$ and inhomogenous linewidth $\Gamma_D$ were taken as free parameters of the fit. Assuming the homogenous linewidth $\Gamma_N\ll\Gamma_D$ (as we observe in our fitting, taking into account possible self-broadening effects at higher densities \cite{Weller2011}), the absorption pre-factor $d^*$ can be related to the "on-resonance" optical depth $d$ via

\begin{equation}
d=\frac{d^*}{\sqrt{\pi \log(2)}}\frac{\Gamma_D}{\Gamma_N}.\label{eq:DstarD}
\end{equation}

The desorbing LIAD beam is derived from an ECDL at 780 nm, which is amplified via a tapered amplifier, and sent into a single mode fibre. With this we generate up to 300 mW of LIAD light which we couple, counter-propagating with respect to the weak probe, into the HC-PCF. The effective intra-fibre LIAD power was around 80 mW. In order to separate the residual back-reflected (mostly from the HC-PCF end face) LIAD beam from the weak probe, we use a combination of polarisation and spectral filtering (Fig. \ref{fig:setup} (c)).

\begin{figure}[h!]
\centerline{\includegraphics[width=\columnwidth]{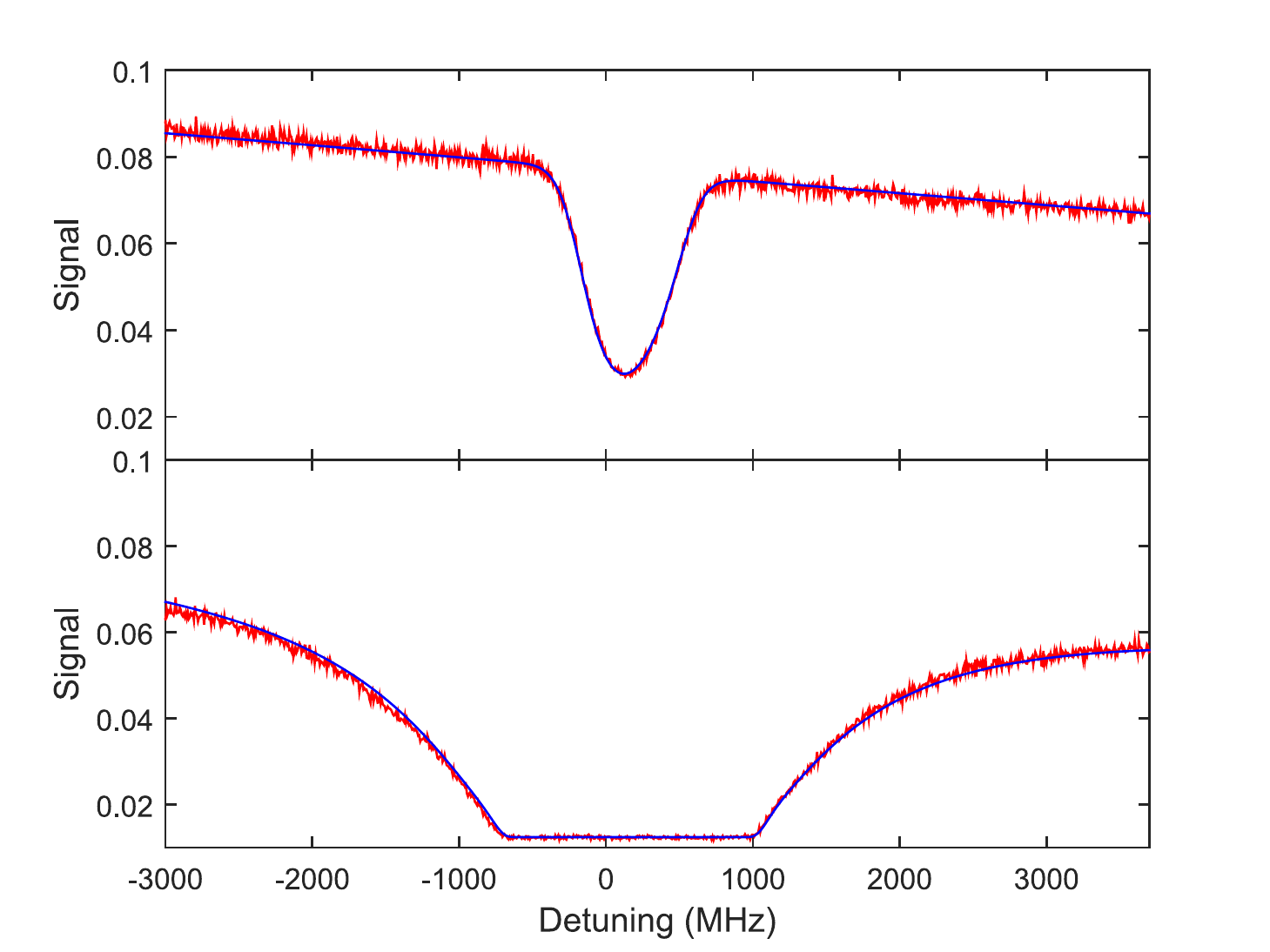}}
\caption{Measured (red) and fitted (blue) transmission spectra of a weak probe beam scanned over the $6S_{1/2}(F=3)\rightarrow6P_{3/2}$ transitions in Cs in a $46\:\mu$m core fibre. Top: no LIAD, fitted $d^*=1.79\pm0.10$, $d=86.6\pm6.1$; bottom: LIAD, fitted $d^*=4290\pm55$, $d=235000\pm5000$.}\label{fig:example}
\end{figure}

In order to explore the long-lived temporal dynamics of the optical depth due to LIAD in our system, we take a series of over 20,000 consecutive probe transmission scans (which correspond to three hours of data acquisition), during which we turn LIAD on and off with an electronically-controlled shutter. Figure \ref{fig:example} shows typical measured transmission, along with fits, of the weak probe beam through the $46\:\mu$m core fibre when its frequency is scanned over the D2 line of Cs for the case of LIAD being off (top) and at the peak OD of it being on (bottom). Due to the large cores of kagom\'e-type HC-PCFs, there is a measurable trace of non-adsorbed alkali in the vapour phase even at room temperature with LIAD being off.

\begin{figure}[h!]
\centerline{\includegraphics[width=\columnwidth]{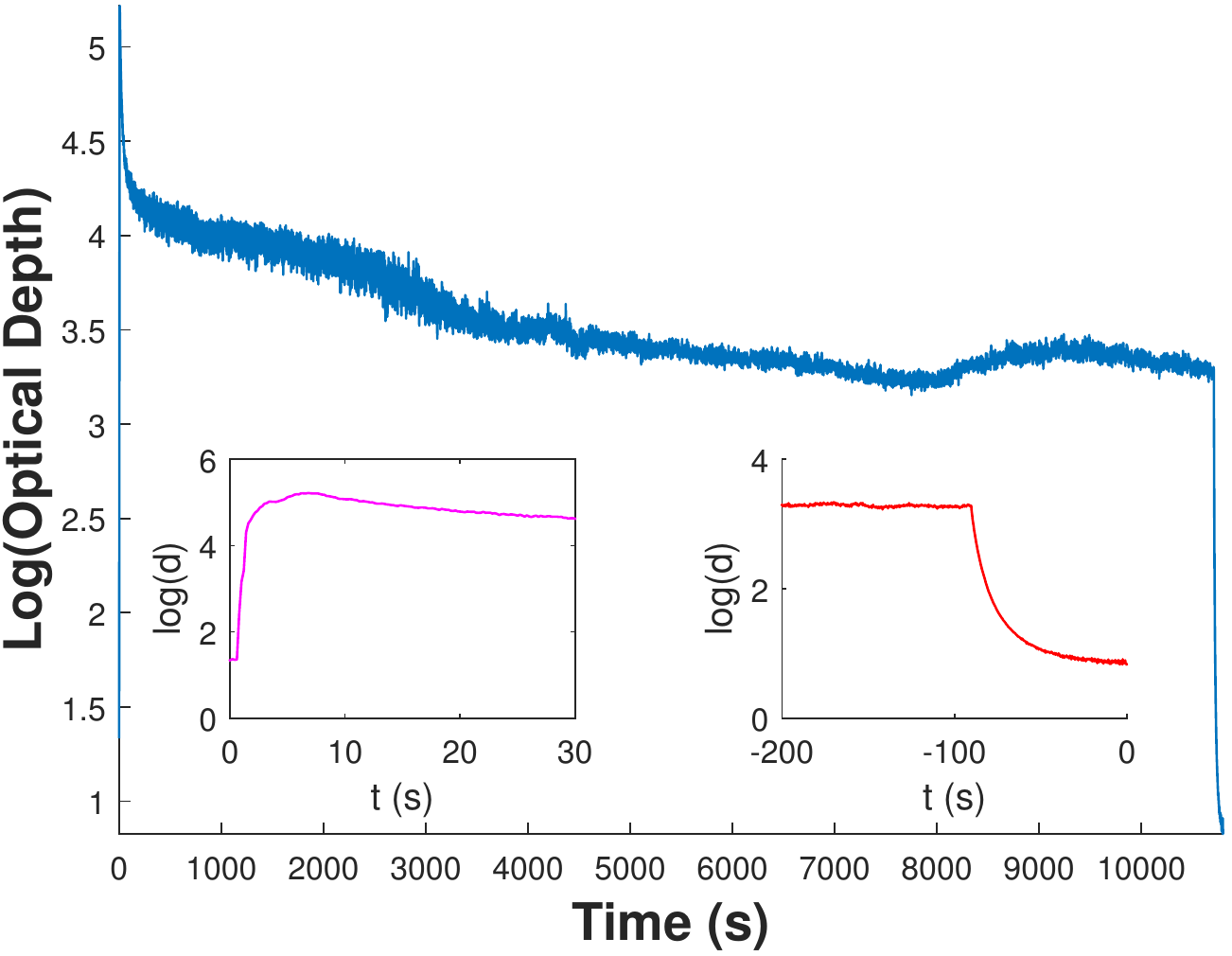}}
\caption{Change in optical depth of Cs over 3 hours of LIAD illumination in a $26\:\mu$m core fibre. Insets show the behaviour of the optical depth at the beginning (left-bottom) and end (right-bottom) of LIAD illumination.}\label{fig:ODlong26}
\end{figure}

Figure \ref{fig:ODlong26} shows the change in optical depth of Cs in a $26\:\mu$m Kagom\'e HC-PCF when LIAD illumination is applied at room temperature over three hours. Within the span of 10s the OD rises by three orders of magnitude and reaches a maximum value of $1.7\times10^5$, which corresponds to an average number density of atoms $7.6\times10^{11}$ cm$^{-3}$\cite{ReimThesis}.

After the sharp initial rise, the OD drops by an order of magnitude over 4 minutes. This is a significantly longer time for which such high ODs have been observed in a similar system before \cite{Sprague2013}. Over the next 3 hours the optical depth remains above 2000. This is the highest persistent LIAD-induced OD ever observed.

\begin{figure}[h!]
\centerline{\includegraphics[width=\columnwidth]{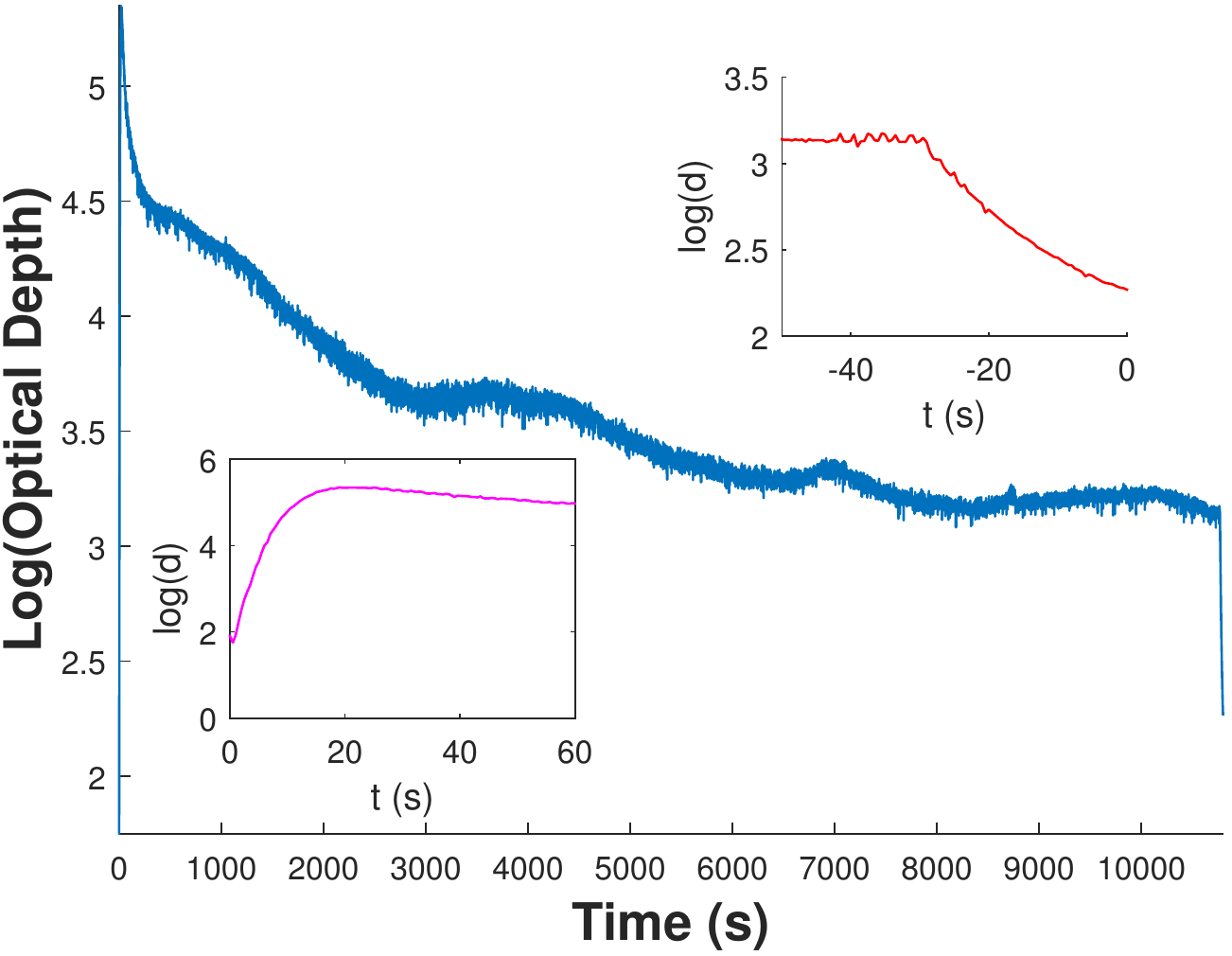}}
\caption{Change in optical depth of Cs over 3 hours of LIAD illumination in a $46\:\mu$m core fibre. Insets show the behaviour of the optical depth at the beginning (left-bottom) and end (right-top) of LIAD illumination.}\label{fig:ODlong44}
\end{figure}

Figure \ref{fig:ODlong44} shows similar data for the $46\:\mu$m core fibre. The initial rise in OD is slightly slower than in the $26\:\mu$m core fibre (20s rise time), but the final OD reached is higher, at $2.2\times10^5$, after which it stays above $10^5$ for up to a minute. This is to our knowledge the highest reached OD in an alkali-filled hollow-core fibre system and is comparable to the ODs generated by heating bulk cells to hundreds of degrees, even though the fibre system operates at room temperature, extending its operational lifetime, and allowing for faster and more precise control of the optical depth.

The visible mid-term fluctuations in LIAD-induced OD can be attributed to changing LIAD powers, due to experimental instabilities. Switching off the LIAD beam (at $t\approx-90$s and $t\approx-30$s in the insets of Fig. \ref{fig:ODlong26} and \ref{fig:ODlong44}, respectively) causes the OD to rapidly drop.

The LIAD-induced changes in OD appear to be governed by two processess of different timescales: a sharp initial rise and fall in OD over the course of a few minutes and a subsequent significantly slower fall in OD over the course of hours. The initial sharp peak in OD only occurs for the first run of the experiment per day and requires around a day of quiescence to regenerate to its former level, which is consistent with previous observations in similar systems \cite{Sprague2013,Bhagwat2009}. This is attributed to the highly efficient light-induced desorption of alkali nanoclusters, which require several hours of quiescence to reform on the surface of the fibre core in order to be efficiently desorbed again. High-power light can also desorb alkalis deposited as metallic layers. Layer desorption is considered to be much less efficient, but leads to a constant elevated level of OD, since the re-adsorbed alkali can be immediately desorbed \cite{Bhagwat2009}.

However, considering only the desorption of nanoclusters and layers, and their reformation/readsorption, does not explain the slow fall in OD observed in our experiment. In order for our model of LIAD to mimic the data correctly, we had to include another process. This process decreases the amount of Cs in the whole system under LIAD illumination, but regenerates it during the period of darkness. In coated bulk cells, the process responsible for similar long-term depletion of desorbable alkali was identified to be its interaction with the coating itself, including light-stimulated absorption into the coating \cite{Atutov1999}. A similar depletion effect was also observed in LIAD from porous alumina, where the pores acted as a sink for the atoms \cite{Villalba2010}, and in vapour cells with closable stems \cite{Karaulanov2009}. We speculate that in glass hollow-core fibres the mechanism responsible for long-term depletion of alkali vapour during LIAD illumination is the escape of the vapour out of the fibre ends, due to a strong, LIAD-induced vapor pressure differential on the boundary between the fibre and its surroundings.
We have devised a rate equation model based on the assumptions outlined above, obtaining the following set of equations:

\begin{equation}
\begin{array}{lcl}
\partial_tN_{\mathrm{v}}&=&R_{\mathrm{l}}^{\mathrm{LIAD}} P_{\mathrm{LIAD}} N_{\mathrm{l}}+R_{\mathrm{c}}^{\mathrm{LIAD}} P_{\mathrm{LIAD}} N_{\mathrm{c}}\\
&&-R^{\mathrm{ads}} (N_{\mathrm{v}}-\mu_{\mathrm{l}} N_{\mathrm{l}} )-R^{\mathrm{esc}} (N_{\mathrm{v}}-\mu_{\mathrm{s}} N_{\mathrm{s}})\\
\partial_tN_{\mathrm{l}}&=&-R_{\mathrm{l}}^{\mathrm{LIAD}} P_{\mathrm{LIAD}} N_{\mathrm{l}}+R^{\mathrm{ads}} (N_{\mathrm{v}}-\mu_{\mathrm{l}} N_{\mathrm{l}} )\\
\partial_tN_{\mathrm{c}}&=&-R_{\mathrm{c}}^{\mathrm{LIAD}} P_{\mathrm{LIAD}} N_{\mathrm{c}}\\
\partial_tN_{\mathrm{s}}&=&R^{\mathrm{esc}} (N_{\mathrm{v}}-\mu_{\mathrm{s}} N_{\mathrm{s}})\\
\end{array}
\label{eq:RateEq}
\end{equation}

\noindent where $N_\mathrm{v},N_\mathrm{l},N_\mathrm{c}$ are the fractions of atoms in the vapour, layer and nanocluster phase, respectively. The first two terms on the right-hand side of the first equation describe LIAD, i.e. the light-induced desorption of metallic layers and nanoclusters. The desorption rates are $R^\mathrm{LIAD}_\mathrm{l}$ and  $R^\mathrm{LIAD}_\mathrm{c}$ for the layers and nanoclusters respectively, with nanocluster desorption much more efficient ($R^\mathrm{LIAD}_\mathrm{c}\gg R^\mathrm{LIAD}_\mathrm{l}$). $P_\mathrm{LIAD}$ is the LIAD beam power. The next two terms describe the equilibrium processes of alkali adsorption as layers and its escape out of the fibre as vapour. $R^\mathrm{ads}$ and $\mu_\mathrm{l}$ are the adsorption rate (fast) and equilibrium ratio of atoms in the vapour and layer phases. $N_\mathrm{s}$ is the fraction of atoms in the sink, i.e. outside the fibre, and $R^\mathrm{esc}$ and $\mu_\mathrm{s}$ are the atomic escape rate from the fibre (slow) and the equilibrium ratio of atoms on the outside and inside the fibre (in the vapour phase).

\begin{figure}[h!]
\centerline{\includegraphics[width=\columnwidth]{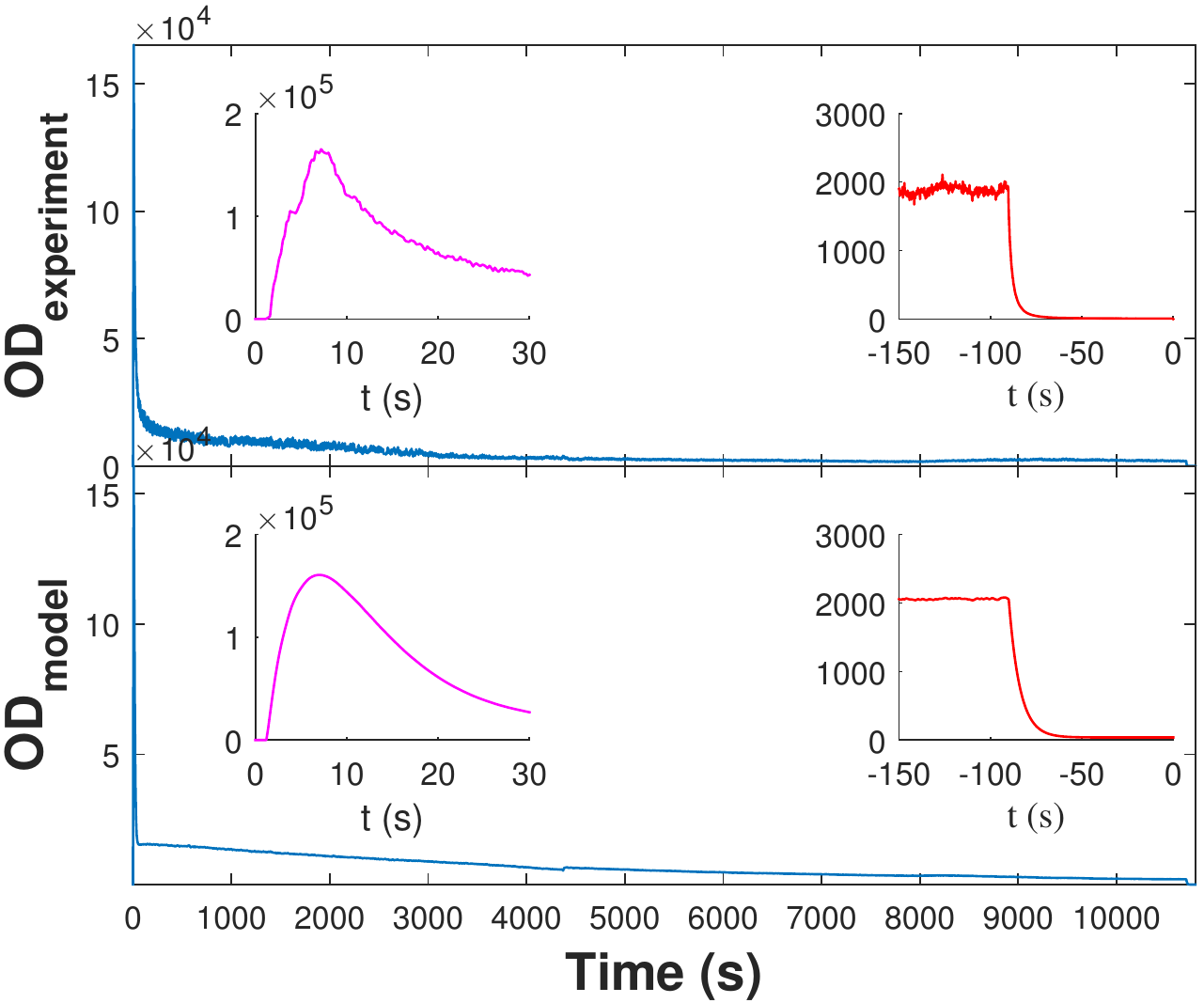}}
\caption{Comparison of data for the $26\:\mu$m core fibre (top) and our rate-equation model (bottom). Insets show the behaviour of the optical depth at the beginning (left) and end (right) of LIAD illumination.}\label{fig:Model}
\end{figure}

Figure \ref{fig:Model} shows a comparison between experimental data (top) and the numerical solution of the model (bottom). The model was fed with the measured LIAD power in the fibre, and the rate coefficients from Eq. \ref{eq:RateEq} were found manually (shown in Table \ref{tab:tab1}), in order to get the closest similarity to data. Indeed, the model accurately describes the qualitative features of the data, confirming that the process limiting the long-term OD is reversible absorption of the atoms into a sink during prolonged LIAD illumination, which physically could be the escape of the atomic vapour into the space outside the fibre ends.

\begin{table}[htbp]
\centering
\caption{\bf Rate coefficients from Eq. \ref{eq:RateEq} used to generate the model shown in Fig. \ref{fig:Model}.}
\begin{tabular}{cccc}
\hline
$R_{\mathrm{c}}^{\mathrm{LIAD}}$ & $R_{\mathrm{l}}^{\mathrm{LIAD}}$&$R^{\mathrm{ads}}$&$R^{\mathrm{esc}}$\\
\hline
6& 0.06&$15\times10^{-2}$& $16\times10^{-7}$\\
\end{tabular}
  \label{tab:tab1}
\end{table}

In summary, we have constructed a caesium-filled hollow-core fibre system which combines high alkali atomic densities with strong light guidance at room temperature. After passively loading two fibres of $26\:\mu$m and $46\:\mu$m core diameters with the alkali for almost a year, we generate the highest observed transient ODs $>10^5$ in this system using light-induced atomic desorption. We are then able to maintain ODs larger than 2000 for at least three hours. We attribute this to a long loading time of the fibre and high in-coupled LIAD optical power. We observe no significant difference in the LIAD dynamics between the two core sizes. Using a rate-equation model approach, we identified the loss of atoms, possibly due to their escape from the fibre during LIAD illumination, as the main process responsible for the long-term fall in OD.

A possible way to prevent the long-term drop in OD is to seal the alkali-filled HC-PCF ends to prevent the atoms inside from escaping. This can be achieved for example, by splicing the HC-PCF to single mode fibres, creating a fibre-based microcell \cite{Benabid2009}, which has the additional benefit of being integratable into current photonic networks. The duration of the transient ultrahigh ODs due to nanocluster desorption would still be limited by the long reformation time of the nanoclusters.

These results open the way towards using an alkali-filled fibre for conducting time- and atomic density-intensive experiments, such as investigating single photon quantum memories \cite{Michelberger2015} or single photon nonlinearities \cite{Venkataraman2013}.

\section*{Acknowledgements}
This work was supported by Air Force Office of Scientific Research (AFOSR)
(FA8655-09-1-3020); Engineering and Physical Sciences
Research Council (EPSRC) (EP/K034480/1 (BLOQS); EP/
J000051/1, NQIT); European Commission (EC) (PIEFGA-2013-62737,
IEF-GA-2012-331859, IP SIQS (600645),
PIIF-GA-2013-629229); European Research Council (ERC)
(MOQUACINO); Royal Society.

We thank A. Abdolvand and P.~St.~J. Russell from the Max Planck Institute for the Science of Light (Germany) for providing us with the hollow-core fibres and their characterization. We would also like to thank I.~G. Hughes for his insighftul comments on the early version of the manuscript.

\noindent Note added: while preparing this manuscript, we became aware of similar results in smaller-core photonic bandgap fibres \cite{Donvalkar2015}.


\newpage
\begin{thebibliography}{10}
\newcommand{\enquote}[1]{``#1''}

\bibitem{Humphreys2014}
P.~C. Humphreys and W.~S. Kolthammer, \enquote{{Continuous-Variable Quantum
  Computing in Optical Time-Frequency Modes using Quantum Memories},} Physical
  Review Letters \textbf{130502}, 5 (2014).

\bibitem{Nunn2013}
J.~Nunn, N.~K. Langford, W.~S. Kolthammer, T.~F.~M. Champion, M.~R. Sprague, P. S. Michelberger, X.-M. Jin, D. G. England, and I. A. Walmsley,
  \enquote{{Enhancing Multiphoton Rates with Quantum Memories},} Physical
  Review Letters \textbf{110}, 133601 (2013).

\bibitem{Benabid2002}
F.~Benabid, J.~C. Knight, G.~Antonopoulos, and P.~S.~J. Russell,
  \enquote{{Stimulated Raman scattering in hydrogen-filled hollow-core photonic
  crystal fiber.}} Science (New York, N.Y.) \textbf{298}, 399--402 (2002).

\bibitem{Ayrin2015}
C. Sayrin, C. Clausen, B. Albrecht, P. Schneeweiss, and A. Rauschenbeutel, \enquote{{Storage of fiber-guided light in a nanofiber-trapped
  ensemble of cold atoms},} Optica \textbf{2}, 353--356 (2015).

\bibitem{Gouraud2015}
B.~Gouraud, D.~Maxein, A.~Nicolas, O.~Morin, and J.~Laurat,
  \enquote{{Demonstration of a Memory for Tightly Guided Light in an Optical
  Nanofiber},} Physical Review Letters \textbf{114}, 1--5 (2015).

\bibitem{Venkataraman2013}
V.~Venkataraman, K.~Saha, and A.~Gaeta, \enquote{{Phase modulation at the
  few-photon level for weak-nonlinearity-based quantum computing},} Nature
  Photonics \textbf{7}, 5--8 (2013).

\bibitem{Epple2014}
G.~Epple, K.~S. Kleinbach, T.~G. Euser, N.~Y. Joly, T.~Pfau, P.~S.~J. Russell,
  and R.~L\"{o}w, \enquote{{Rydberg atoms in hollow-core photonic crystal
  fibres.}} Nature communications \textbf{5}, 4132 (2014).

\bibitem{Sprague2014}
M.~R. Sprague, P.~S. Michelberger, T.~F.~M. Champion, D.~G. England, J.~Nunn,
  X.-M. Jin, W.~S. Kolthammer, A.~Abdolvand, P.~S.~J. Russell, and I.~A.
  Walmsley, \enquote{{Broadband single-photon-level memory in a hollow-core
  photonic crystal fibre},} Nature Photonics \textbf{8}, 287--291 (2014).

\bibitem{Bhagwat2009}
A.~Bhagwat, A.~Slepkov, V.~Venkataraman, P.~Londero, and A.~Gaeta,
  \enquote{{On-demand all-optical generation of controlled Rb-vapor densities
  in photonic-band-gap fibers},} Physical Review A \textbf{79}, 063809 (2009).

\bibitem{Vogl2014}
U.~Vogl, C.~Peuntinger, N.~Y. Joly, P.~S. Russell, C.~Marquardt, and G.~Leuchs,
  \enquote{{Atomic mercury vapor inside a hollow-core photonic crystal fiber},}
  Optics Express \textbf{22}, 29375 (2014).

\bibitem{Blatt2014}
F.~Blatt, T.~Halfmann, and T.~Peters, \enquote{{One-dimensional ultracold
  medium of extreme optical depth},} Optics Letters \textbf{39}, 446 (2014).

\bibitem{Slepkov2008}
A.~D. Slepkov, A.~R. Bhagwat, V.~Venkataraman, P.~Londero, and A.~L. Gaeta,
  \enquote{{Generation of large alkali vapor densities inside bare hollow-core
  photonic band-gap fibers.}} Optics express \textbf{16}, 18976--83 (2008).

\bibitem{Burchianti2008}
A.~Burchianti, A.~Bogi, C.~Marinelli, C.~Maibohm, E.~Mariotti, S.~Sanguinetti,
  and L.~Moi, \enquote{{Optical characterization and manipulation of alkali
  metal nanoparticles in porous silica},} The European Physical Journal D
  \textbf{49}, 201--210 (2008).

\bibitem{Burchianti2010}
A.~Burchianti, A.~Bogi, C.~Marinelli, E.~Mariotti, and L.~Moi,
  \enquote{{Optical stabilization of Rb vapor density above thermal
  equilibrium},} Journal of Modern Optics \textbf{57}, 1305--1310 (2010).

\bibitem{Weller2011}
L.~Weller, R.~J. Bettles, P.~Siddons, C.~S. Adams, and I.~G. Hughes,
  \enquote{{Absolute absorption on the rubidium D1 line including resonant
  dipole–dipole interactions.}} Journal of Physics B: Atomic, Molecular and
  Optical Physics \textbf{195006}, 0--5 (2011).

\bibitem{ReimThesis}
K. F. Reim, \enquote{Broadband optical quantum memory}, DPhil thesis, University of Oxford (2011).

\bibitem{Sprague2013}
M.~R. Sprague, D.~G. England, A.~Abdolvand, J.~Nunn, X.-M. Jin, W.~S.
  Kolthammer, M.~Barbieri, B.~Rigal, P.~S. Michelberger, T.~F.~M. Champion,
  P.~S.~J. Russell, and I.~A. Walmsley, \enquote{{Efficient optical pumping and
  high optical depth in a hollow-core photonic-crystal fibre for a broadband
  quantum memory},} New Journal of Physics \textbf{15}, 055013 (2013).

\bibitem{Atutov1999}
S.~Atutov, V.~Biancalana, and P.~Bicchi, \enquote{{Light-induced diffusion and
  desorption of alkali metals in a siloxane film: Theory and experiment},}
  Physical Review A \textbf{60}, 4693--4700 (1999).

\bibitem{Villalba2010}
S.~Villalba, H.~Failache, and A.~Lezama, \enquote{{Light-induced atomic
  desorption and diffusion of Rb from porous alumina},} Physical Review A
  \textbf{81}, 032901 (2010).

\bibitem{Karaulanov2009}
T.~Karaulanov, M.~T. Graf, D.~English, S.~M. Rochester, Y.~J. Rosen,
  K.~Tsigutkin, D.~Budker, E.~B. Alexandrov, M.~V. Balabas, D.~F.~J. Kimball,
  F.~A. Narducci, S.~Pustelny, and V.~V. Yashchuk, \enquote{{Controlling atomic
  vapor density in paraffin-coated cells using light-induced atomic
  desorption},} Physical Review A - Atomic, Molecular, and Optical Physics
  \textbf{79}, 1--9 (2009).

\bibitem{Benabid2009}
F.~Benabid, P.~J. Roberts, F.~Couny, and P.~S. Light, \enquote{{Light and gas
  confinement in hollow-core photonic crystal fibre based photonic
  microcells},} Journal of the European Optical Society: Rapid Publications
  \textbf{4}, 09004 (2009).

\bibitem{Michelberger2015}
P.~S. Michelberger, T.~F.~M. Champion, M.~R. Sprague, K.~T. Kaczmarek,
  M.~Barbieri, X.~M. Jin, D.~G. England, W.~S. Kolthammer, D.~J. Saunders,
  J.~Nunn, and I.~A. Walmsley, \enquote{{Interfacing GHz-bandwidth heralded
  single photons with a room-temperature Raman quantum memory},} New Journal of
  Physics \textbf{17}, 43006 (2015).

\bibitem{Donvalkar2015}
P.~S. Donvalkar, S.~Ramelow, S.~Clemmen, and A.~L. Gaeta, \enquote{{Continuous
  generation of Rubidium vapor in hollow-core photonic band-gap fibers},} arXiv:1509.04737 [physics.atom-ph] (2015).



\end{thebibliography}
\end{document}